\documentclass[12pt]{article}
\usepackage{amsmath}

 \newcommand{\lan}{\langle}
 \newcommand{\ran}{\rangle}
 \newcommand{\be}{\begin{equation}}
 \newcommand{\bea}{\begin{eqnarray}}
 \newcommand{\eea}{\end{eqnarray}}
 \newcommand{\ee}{\end{equation}}

\def\la{\mathrel{\mathpalette\fun <}}

\def\fun#1#2{\lower3.6pt\vbox{\baselineskip0pt\lineskip.9pt
\ialign{$\mathsurround=0pt#1\hfil##\hfil$\crcr#2\crcr\sim\crcr}}}
\newcommand{\vep}{\mbox{\boldmath${\rm p}$}}

\title{Low temperature relations in QCD}
\author{
Nikita~O.~Agasian\thanks{e-mail: agasian@heron.itep.ru} \\
{\it Institute of Theoretical and Experimental Physics} \\
{\it 117218, Moscow, Russia}}
\date{}
\begin{document}
\maketitle \vspace{1cm} {\centerline {\bf Abstract}}
In this talk
I discuss the low temperature relations for the trace of the
energy-momentum tensor in QCD with two  and three quarks. It is
shown that the temperature derivatives of the anomalous and normal
(quark massive term) contributions to the trace of the
energy-momentum tensor in  QCD  are equal to each other in the low
temperature region. Leading corrections connected with $\pi\pi$-interactions and
thermal excitations of $K$ and $\eta$ mesons are calculated.

\vspace{1cm}

%\pacs{11.10.Wx,12.38.Aw,12.38.Mh}

\maketitle
%\newpage

 The investigation of the vacuum state behavior under the
 influence of various external factors is known to be one of the
 central problems of quantum field theory. In the realm of strong
 interactions (QCD) the main factors are the temperature and
 the baryon density. At low temperatures, $T<T_c$ ( $T_c$--temperature of the
 "hadron--quark-gluon" phase transition ), the dynamics of QCD
 is essentially nonperturbative and
 is characterized by confinement and spontaneous
breaking of chiral symmetry (SBCS). In the hadronic phase the
partition function of the system is dominated by the contribution
 of the lightest  particles in the physical spectrum.
It is well known that due to the smallness of pion mass as
compared to the typical scale of strong interactions, the pion
plays a special role among other strongly-interacting particles.
Therefore for many problems of QCD at zero temperature the chiral
limit, $M_\pi\to 0$, is an appropriate one. On the other hand a
new mass scale emerges in the physics of QCD phase transitions,
namely the critical transition temperature $T_c$.
Numerically the critical transition temperature turns out to be
close to the pion mass, $T_c\approx M_\pi$ \footnote{
The deconfining phase transition temperature is the  one obtained in
lattice calculations $T_c(N_f=2)\simeq 173 $ MeV and
$T_c(N_f=3)\simeq  154$ MeV \cite{kar}.}.
However hadron states heavier than pion have masses several
times larger than $T_c$ and therefore their contribution
to the thermodynamic quantities is damped
by Boltzmann factor $\sim \exp \{-M_{hadr}/T\}$.
Thus the thermodynamics of the low temperature
hadron phase, $T\la M_\pi$, is described basically in terms of the
thermal  excitations of relativistic massive pions.

 The low-energy theorems, playing an important role in the understanding
 of the vacuum state properties in quantum field theory,
were discovered almost at the same time as quantum field methods have been applied
in particle physics (see, for example, Low theorems \cite{low}).
In QCD, they were obtained in the beginning of eighties
\cite{LT}. The QCD low-energy theorems, being derived from the very general
symmetry considerations and not depending on the details
of confinement mechanism, sometimes give information which
is not easy to obtain in another way. Also, they can be used
as "physically sensible" restrictions in the constructing
of effective theories. Recently, they were generalized to
finite temperature \cite{k1} and a magnetic field case \cite{a1}.
These theorems were used for investigation of QCD vacuum phase
structure in a magnetic field at finite
temperature \cite{a2}.

In this talk I will discuss the low temperature relations
for the trace of the energy-momentum tensor in QCD
with two  and three light quarks. These relations
is based on the general dimensional and
renormalization-group properties of the QCD partition function
and dominating role of the pion thermal exitations in the hadronic phase.
The physical consequences of these relations are discussed
as well as the possibilities to use it in the lattice studies
of the QCD at finite temperature.

 For non-zero quark mass ($m_q\neq 0)$ the scale invariance is
 broken already at the classical level.
Therefore the pion thermal  excitations would change, even in the
ideal gas approximation, the value of the gluon condensate
with increasing temperature \footnote{
At zero quark mass
 the gas of massless noninteracting pions is obviously
 scale-invariant and therefore does not contribute to
the trace of the  energy-momentum tensor and correspondingly to the
gluon condensate $\langle G^2\rangle \equiv \langle (g G^a_{\mu\nu})^2\rangle$.}.

 A relation between the trace anomaly and thermodynamic pressure in the chiral
 limit of QCD was first written in~\cite{9}. Rigorous derivation of this relation
in the framework of the renormalization group (RG) method in pure-glue QCD was
performed in \cite{Land} and in QCD with non-zero quark masses in~\cite{k2,agjetp}.

Trace anomaly is~\cite{agjetp}

 \be
\lan \theta^g_{\mu\mu}\ran
=\frac{\beta(\alpha_s)}{16\pi\alpha_s^2} \langle G^2\rangle
=-(4-T\frac{\partial}{\partial
T}-\sum_q(1+\gamma_{m_q})m_q\frac{\partial}{\partial {m_q}}) P_R,
\label{10a}
 \ee
 where
 $P_R$ is the renormalized pressure, $\beta(\alpha_s)=d\alpha_s(M)/d
  ~ln M$  is the Gell-Mann-Low  function and $\gamma_{m_q}$ is the
anomalous dimension of the quark mass.
   It is convenient to choose such a large scale that one can take the
lowest order expressions,  $\beta(\alpha_s)\to -
b\alpha^2_s/2\pi$, where $b=(11 N_c-2N_f)/3$ and $1+\gamma_m\to
1$. Thus, we have the following equations for condensates
\be
\lan G^2\ran (T)=\frac{32\pi^2}{b} (4-T\frac{\partial}{\partial
T}-\sum_q m_q\frac{\partial}{\partial m_q}) P_R\equiv
 \hat DP_R~,
 \label{11}
 \ee
 \be
 \lan\bar q q\ran (T)=-\frac{\partial P_R}{\partial {m_q}}~.
 \label{12}
 \ee

  In the hadronic phase the effective pressure from which one can extract the
condensates $\lan \bar q q\ran(T)$ and $\lan G^2\ran(T)$ using
the general relations (\ref{11}) and (\ref{12}) has the form
\be
P_{eff}(T)=-\varepsilon_{vac}+P_h(T),
\label{14}
\ee
where
$\varepsilon_{vac}=\frac14\lan\theta_{\mu\mu}\ran$ is the
nonperturbative vacuum energy density at $T=0$ and
\be
\lan
\theta_{\mu\mu}\ran=-\frac{b}{32\pi^2} \lan G^2\ran+\sum_{q=u,d}
m_q\lan\bar qq \ran
\label{15}
\ee
is the trace of the
energy-momentum tensor. In Eq.(\ref{14}) $P_h(T)$ is the
thermal hadrons pressure.
The quark and gluon condensates are given by the equations
\be
\lan \bar qq\ran
(T)=-\frac{\partial P_{eff}}{\partial m_q},
\label{17}
\ee
\be
\lan G^2\ran (T)= \hat DP_{eff},
\label{18}
\ee
where the
operator $\hat D$ is defined by the relation (\ref{11})
\be \hat
D=\frac{32\pi^2}{b} (4-T\frac{\partial}{\partial T}-\sum_q
m_q\frac{\partial}{\partial m_q})~.
\label{19}
\ee

Consider the $T=0$ case. One can use the low energy theorem
for the derivative of the gluon condensate with respect to the
quark mass \cite{LT}
\be
\frac{\partial}{\partial m_q}\lan G^2\ran= \int d^4 x\lan G^2(0)
\bar q q(x)\ran =-\frac{96\pi^2}{b}\lan \bar q q\ran+O(m_q),
\label{20}
\ee
where $O(m_q)$ stands for the terms linear in light
quark masses.Then one arrives at the following
relation
\be
\frac{\partial\varepsilon_{vac}}{\partial m_q}=-
\frac{b}{128\pi^2}\frac{\partial}{\partial m_q} \lan
G^2\ran+\frac{1}{4}\lan \bar q q\ran =\frac34 \lan \bar q
q\ran+\frac14\lan \bar q q\ran=\lan \bar q q\ran.
\label{21}
\ee
Note that three fourths of the quark condensate stem from the
gluon part of the nonperturbative vacuum energy density. Along the
same lines one arrives at the expression for the gluon
condensate
\be
-\hat D\varepsilon_{vac}=\lan G^2\ran.
\label{22}
\ee

Let us consider $N_f=2$ case. In order to get the dependence of the quark and
gluon condensates upon $T$ use is made of the Gell-Mann-
Oakes-Renner (GMOR) relation ($\Sigma=|\lan\bar u u\ran|=|\lan
\bar dd\ran|$)
\be F^2_\pi M^2_\pi=-\frac12(m_u+m_d)\lan \bar
uu+\bar dd\ran=(m_u+m_d)\Sigma~.
\label{23}
\ee
Then we can find the following relations
\be
\frac{\partial}{\partial
m_q}=\frac{\Sigma}{F^2_\pi} \frac{\partial}{\partial M^2_\pi}~,
\label{24}
\ee
\be \sum_{q=u,d} m_q\frac{\partial}{\partial
m_q}=(m_u+m_d)\frac{\Sigma}{F^2_\pi}\frac{\partial}{\partial
M^2_\pi}=M^2_\pi\frac{\partial}{\partial M^2_\pi}~,
\label{25}
\ee
\be \hat D=\frac{32\pi^2}{b}(4-T\frac{\partial}{\partial
T}-M^2_\pi\frac{\partial}{\partial M^2_\pi})~.
\label{26}
\ee

 At low temperature the main
 contribution to the pressure comes from thermal excitations of
 massive pions. The general expression for the pressure reads
 \be
 P_\pi=T^4\varphi(M_\pi/T)~,
 \label{36}
 \ee
 where $\varphi$ is a function of the ratio $M_\pi/T$.
 Then the following  relation is valid
 \be
 (4-T\frac{\partial}{\partial T}-M^2_\pi\frac{\partial}{\partial
 M^2_\pi}) P_\pi=M^2_\pi\frac{\partial P_\pi}{\partial M^2_\pi}~.
 \label{37}
 \ee
With the account of (\ref{17},\ref{18}), (\ref{21},{22}) and (\ref{37})
one gets
\be
\Delta \lan \bar qq\ran =-\frac{\partial P_\pi}{\partial m_q},~~
\Delta \lan G^2\ran=\frac{32\pi^2}{b} M^2_\pi\frac{\partial
P_\pi}{\partial M^2_\pi}~,
\label{38}
\ee
where $ \Delta \lan \bar
qq\ran= \lan \bar qq\ran_T- \lan \bar qq\ran $ and $\Delta \lan
G^2\ran=  \lan G^2\ran_T- \lan G^2\ran.$ In view of (\ref{25}) one
can recast (\ref{38}) in the form
\be
\Delta \lan G^2\ran=-\frac{32\pi^2}{b} \sum_{q=u,d} m_q\Delta  \lan \bar
qq\ran~.
\label{39}
\ee
Differentiating~(\ref{39}) with respect to $T$ one obtains
\be
\frac{\partial \lan G^2\ran}{\partial T}=-\frac{32\pi^2}{b} \sum_{q=u,d}
m_q\frac{\partial\lan \bar qq\ran}{\partial T}~.
\label{40}
\ee
This can be rewritten as \cite{agplb}
\be
\frac{\partial \lan \theta^g_{\mu\mu}\ran}{\partial
T}=\frac{\partial \lan \theta^q_{\mu\mu}\ran}{\partial T}~,
\label{41}
\ee
where $\lan \theta^q_{\mu\mu}\ran=\sum m_q \lan
\bar qq\ran$ and $\lan
\theta^g_{\mu\mu}\ran=(\beta(\alpha_s)/16\pi\alpha^2_s) \lan
G^2\ran$ are correspondingly the quark and gluon contributions
to the trace of the energy-momentum tensor.
Note that in deriving this result use was made of
the low energy GMOR relation, and therefore the thermodynamic
relation (\ref{40},\ref{41}) is valid in the light quark theory.
Thus in the low temperature region when the excitations of massive
hadrons and interactions of pions can be neglected, equation
(\ref{41}) becomes a rigorous QCD theorem.

As it was mentioned above the pion plays an exceptional role in
thermodynamics of QCD due to the fact that its mass is numerically
close to the phase transition temperature while the masses of
heavier hadrons are several times larger than $T_c$.
It was shown in \cite{GL} that at low temperatures, the
contribution to $\lan \bar q q\ran$ generated by the massive
states is very small, less than 5\% if $T$ is below 100 MeV. At
$T=150$ MeV, this contribution is of the order of 10\%. The
influence of thermal excitations of massive hadrons on the
properties of the gluon and quark condensates in the framework of
the conformal-nonlinear $\sigma$- model was studied in detail
in \cite{AEI}.

Let us consider leading corrections to relation~(\ref{39}-\ref{41}) within
the described above framework. Clearly, leading corrections are connected with
the $\pi\pi$-interaction, since it's contribution to the pressure is $\propto e^{-2M_{\pi}/T}$.
Also, account for $s$-quark leads to the contributions
to the pressure $\propto e^{-M_{K}/T}$, $e^{-M_{\eta}/T}$, which are related to
the thermal excitations of $K$ and $\eta$-mesons.
Then pressure in hadronic phase can be recast in the following form

\begin{equation}
P_h(T)=P_g(T)+P_{\pi\pi}(T),
\end{equation}
\begin{equation}
P_g(T)=\sum_{i=\pi,K,\eta} P_{i}(T),
\end{equation}
where $P_{i}(T)=g_{i} T^{4} \varphi(M_{i}/T)$ is gas pressure of $i=\pi,K,\eta$ - meson and
$g_i$ is the number of degrees of freedom of $i$-state, $g_{\pi}=3$, $g_K$=4, $g_{\eta}=1$.
Pressure related to $\pi\pi$ interaction in two-loop ChPT in general form is

\begin{equation}
P_{\pi\pi}=T^4 \frac{M_{\pi}^2}{F_{\pi}^2} f\left(\frac{M_{\pi}}{T}\right),
\label{eq_P_pi_pi}
\end{equation}
here $f$ is function of ratio $M_{\pi}/T$, and factor $M_{\pi}^2/F_{\pi}^2$ is
connected with the $\pi\pi$ interaction vertex.

Making use of Gell-Mann-Okubo relations one gets (analogous to~(\ref{37}))

\begin{align}
\hat D P_g(T)&=\frac{32\pi^2}{b} (4-T\frac{\partial}{\partial T}-
\sum_{q=u,d,s} m_q\frac{\partial}{\partial m_q}) P_g(T) \\
&=\frac{32\pi^2}{b} \sum_{i=\pi,K,\eta} M_i^2 \frac{\partial P_i}{\partial M_i^2} \notag
\end{align}
For the temperature shift of quark condensates on has

\begin{equation}
\frac{\Delta \Sigma(T)}{\Sigma} = \frac{\partial P_g}{\partial m_u} = \frac{1}{F_{\pi}^2}
\left( \frac{\partial P_{\pi}}{\partial M_{\pi}^2} +
       \frac{\partial P_{K}}{\partial M_{K}^2} +
       \frac{1}{3}\frac{\partial P_{\eta}}{\partial M_{\eta}^2} \right)
\end{equation}

\begin{equation}
\frac{\Delta \Sigma_s(T)}{\Sigma_s} = \frac{\partial P_g}{\partial m_s} = \frac{1}{F_{\pi}^2}
\left( \frac{\partial P_{K}}{\partial M_{K}^2} +
       \frac{4}{3}\frac{\partial P_{\eta}}{\partial M_{\eta}^2} \right)
\label{eq_delta_sigma_s}
\end{equation}
Note, that light $\pi$-meson does not carry strangeness and thus does not participate in
$\langle \bar s s \rangle$ condensate ''evaporation''. Leading contribution to
$\Delta \langle \bar s s \rangle (T)$ comes from thermal excitations of lightest strange $K$-meson
with the mass several times larger than $M_{\pi}$. Therefore it is obvious that
$\langle \bar s s \rangle (T)$ decreases more slowly than $\langle \bar u u \rangle =
\langle \bar d d \rangle$ with the increase of $T$. In the gas approximation one finds
\footnote{Contribution of $K$-meson to the $\Delta \Sigma_s(T)$ can be obtained from low
temperature expression
for the condensate $\Delta\Sigma(T)$ (see~\cite{agplb}), with the obvious substitution of
$M_{\pi} \to M_K$, $F_{\pi} \to F_K$ and multiplication by the factor $4/3$.}

\begin{equation}
\frac{\Delta \Sigma_s(T)/\Sigma_s}{\Delta \Sigma(T)/\Sigma} =
\frac{4}{3} \left( \frac{M_K}{M_{\pi}} \right)^{1/2} \left( \frac{F_{\pi}}{F_K} \right)^2
e^{(M_{\pi}-M_K)/T}
\end{equation}
and this ratio is of order of $\sim 0.13$ at $T\sim 140$~MeV. Analogous to the derivation of
equation~(\ref{39}) (with the account of the~(\ref{eq_P_pi_pi}-\ref{eq_delta_sigma_s}))one gets

\be
-\frac{b}{32\pi^2} \Delta \lan G^2\ran=\sum_{q=u,d,s} m_q\Delta \langle \bar q q\rangle
 + 2 P_{\pi\pi} + \frac{1}{2} M_{\pi}^2 \frac{\partial P_K}{\partial M_K^2}
\ee
Let us introduce functions
\begin{equation}
  \theta^{\pm}(T)=\langle \theta^g_{\mu\mu} \pm \theta^{q}_{\mu\mu} \rangle (T) -
                  \langle \theta^g_{\mu\mu} \pm \theta^{q}_{\mu\mu} \rangle (0).
\end{equation}
$\theta^{+}(T)$ is the thermal part of the trace of the energy-momentum tensor and
$\theta^{+}(T) = \Delta \langle \theta^{tot}_{\mu\mu}\rangle(T)=\varepsilon - 3 P$,
where $\varepsilon = T dP/dT - P$ is energy density.

Then the function
\begin{equation}
\delta_{\theta}(T)=\frac{\theta^{-}(T)}{\theta^{+}(T)}=
\frac{2 P_{\pi\pi} + \frac{1}{2} M_{\pi}^2 \frac{\partial P_K}{\partial M_K^2}}{\varepsilon-3 P}
\end{equation}
can be considered as a measure of the deviation from low temperature relation~(\ref{39}). Let us
estimate this correction numerically. One has for $P_{\pi\pi}$~\cite{5}.

\begin{align}
P_{\pi\pi}&=-\frac{1}{6} \left(\frac{M_{\pi}^2}{F_{\pi}^2}\right)
\left[ g_1 (M_{\pi}/T)\right]^2\\
g_1&=\int \frac{d^3p}{(2\pi)^3} \frac{1}{\omega_{\pi}\left( e^{\omega_{\pi}/T}-1\right)}, \quad
\omega_{\pi}=\sqrt{\vep^2 + M_{\pi}^2}
\end{align}

For $i$-meson gas
\begin{equation}
P_i=-g_i T\int\frac{d^3p}{(2\pi)^3}\ln
(1-e^{-\sqrt{\vep^2+M^2_i}/T}).
\end{equation}

Choosing $M_{\pi}=140$~MeV, $M_K=493$~MeV, $F_{\pi}=93$~MeV, one can see from
numerical calculations that

\begin{equation}
\delta_{\theta}(T<150~\mbox{MeV}) < 0.04
\end{equation}
Consequently, leading corrections to the low temperature relation~(\ref{39}) amounts
to several percent up to the critical temperature.

Thus the function $\delta_\theta(T)$ at low
temperatures is, with  good accuracy, close to zero. In the
vicinity and at the phase transition point, i.e. in the region of
nonperturbative vacuum  reconstruction this function changes
drastically. To see it, we first consider pure gluodynamics. It
was shown in \cite{dilat} using the effective dilaton Lagrangian,
that gluon condensate decreases very weakly with the increase of
temperature, up to phase transition point. This result is
physically transparent and is the consequence of Boltzmann
suppression of thermal glueball excitations in the confining
phase.

Further, the dynamical picture of deconfinement
based on the reconstruction of the nonperturbative
gluonic vacuum was suggested in \cite{sim}. Namely, confining and deconfining phases
%according to
%\cite{sim}
differ first of all in the vacuum fields, i.e., in the value of
the gluon condensate and in the gluonic field correlators.
%It was
%argued in \cite{sim} that
The color-magnetic (CM) correlators and their
contribution to the condensate are kept intact across the temperature phase
transition, while the confining color-electric (CE) part abruptly disappears
above $T_c$. Furthermore, there exist numerical lattice measurements
of field correlators near the critical transition temperature $T_c$,
performed by the Pisa group \cite{dig1}, where both CE and CM correlators
are found with good accuracy. These data clearly demonstrate
the strong suppression of CE  component above $T_c$ and persistence
of CM component.
Thus, the function $\delta'_\theta(T)$ can be presented as a
$\delta$-function smeared around the critical point $T_c$ with the
width $\sim \Delta T$ which defines the fluctuation region of phase
transition.

Similar, but more complicated and interesting situation takes
place in the theory with quarks. The function $\delta_\theta(T)$
contains the quark term, proportional to the chiral phase
transition order parameter $\langle \bar qq\rangle (T)$. So it is
interesting to check the relation (\ref{41}) and to study the
behavior of the function $\delta_\theta(T)$  in the lattice QCD
at finite  temperature. It would allow both to test the
nonperturbative QCD vacuum at the low temperatures in the
confining phase and to extract additional information on the
thermal phase transitions in QCD.

\begin{center}

{\bf ACKNOWLEDGMENTS} \\
\end{center}

I am grateful to A.B. Kaidalov, V.A. Novikov, V.A. Rubakov, Yu.A. Simonov
and A.I. Vainshtein for discussions and comments.
The financial support of RFFI grant 00-02-17836 and
INTAS grant N 110 is gratefully acknowledged.

\end{document}